# Comparison of Helioseismic Far-side Active Region Detections with STEREO Far-Side EUV Observations of Solar Activity

**P. C. Liewer**[1] · **J. Qiu**[2] · **C. Lindsey**[3]

**Abstract** Seismic maps of the Sun's far hemisphere, computed from Doppler data from the *Helioseismic and Magnetic Imager* (HMI) on board the *Solar Dynamics Observatory* (SDO) are now being used routinely to detect strong magnetic regions on the far side of the Sun (http://jsoc.stanford.edu/data/farside/). To test the reliability of this technique, the helioseismically inferred active region detections are compared with far-side observation of solar activity from the *Solar TErrestrial RElations Observatory* (STEREO), using brightness in extreme ultraviolet light (EUV) as a proxy for magnetic fields. Two approaches are used to analyze nine months of STEREO and HMI data. In the first approach, we determine whether or not new large east-limb active regions are detected seismically on the far side before they appear Earth side and study how the detectability of these regions relates to their EUV intensity. We find that, while there is a range of EUV intensities for which far-side regions may or may not be detected seismically, there appears to be an intensity level above which they are almost always detected and an intensity level below which they are never detected. In the second approach, we analyze concurrent extreme ultraviolet and helioseismic far-side observations. We find that 100% (22) of the far-side seismic regions correspond to an extreme ultraviolet plage; 95% of these either became a NOAA-designated magnetic region when reaching the east limb or were one before crossing to the far side. A low but significant correlation is found between the seismic signature strength and the EUV intensity of a far-side region.

[1] Jet Propulsion Laboratory, California Institute of Technology, Pasadena, CA 91109, USA (email: Paulett.Liewer@jpl.nasa.gov)
[2] Montana State University, Bozeman, MT 59717, USA
[3] Northwest Research Associates, Boulder, CO 80301, USA

# 1. Introduction

The turbulent solar convection zone is dominated by acoustic waves with periods between two and ten minutes; helioseismology uses these acoustic waves to probe the interior of the Sun. Braun, Duvall and LaBonte (1988) demonstrated how helioseismology could be used to study compact local anomalies of the Sun such as sunspots (see also Braun *et al*., 1992). Lindsey and Braun (1997) developed the technique of acoustic holography which reconstructs the acoustic field in the solar interior using helioseismic observations at the solar surface. It was shown that, using this technique, strong magnetic regions on the far side of the Sun could be imaged by measuring the reductions in the travel time of acoustic waves passing through the Sun from the far-side surface that are detected on the Earth-side hemisphere (Lindsey and Braun, 2000; Braun and Lindsey, 2001). With this technique, far-side active region maps have now been produced using Doppler data from the *Michelson Doppler Imager* (MDI) (Scherrer *et al.*, 1995) on board the *Solar and Heliospheric Observatory* (SOHO); from the National Solar Observatory Global Oscillation Network Group (GONG) (Harvey *et al.*, 1996); and from the *Helioseismic and Magnetic Imager* (HMI) (Schou *et al.*, 2011) on board the *Solar Dynamics Observatory* (SDO). A calibration was derived by González Hernández, Hill, and Lindsey (2007) to relate the measured reduction in the transit time to the far-side active region's magnetic field strength. This calibration was recently confirmed by a direct comparison of Earth-side seismic and magnetic signatures (MacDonald *et al.,* 2015). Carrington maps showing detections of far-side strong field regions are now routinely produced from both HMI and GONG Doppler data; these maps can be found at http://jsoc.stanford.edu/data/farside/ and http://farside.nso.edu/. Because this technique has the potential to observe the emergence and growth of new large active regions on the far hemisphere of the Sun prior to their appearance on the Earth-side hemisphere using only Earth-side observations, it is of great interest to the space weather community. Moreover, space weather predictions utilizing heliospheric models will be improved if reliable detections of strong magnetic fields on the far side of the Sun can be made and incorporated into the models' solar magnetic field boundary conditions as demonstrated by Arge *et al.* (2013).

Beginning in February 2011, the twin *Solar TErrestrial RElations Observatory* (STEREO) spacecraft provided a view of the entire far side of the Sun. This allowed, for the first time, a direct comparison of far-side helioseismic detections of active regions with simultaneous far-side solar activity as observed by the *Extreme Ultraviolet Imager* (EUVI), one of the telescopes in STEREO's *Sun Earth Connection Coronal and Heliospheric Investigation* (SECCHI) imaging suite (Howard *et al.*, 2008; Wuelsler *et al.*, 2004). Liewer *et al.* (2012, 2014) investigated the reliability of the helioseismic far-side active region predictions from both *Global Oscillation Network Group* (GONG) and HMI using a qualitative comparison with STEREO's far-side extreme ultraviolet (EUV) images, where brightness in EUV was used as a proxy for magnetic fields. Comparing the seismic Carrington maps created *via* the data pipeline at the National Solar Observatory with EUV Carrington maps, Liewer *et al.* (2012, 2014, Papers I and II hereafter) determined whether or not excess EUV brightness was observed at the far-side locations predicted in the seismic maps and whether or not new large active regions were predicted before they appeared Earth side. In Paper II, it was found that nearly all (90%) of the far-side seismic detections from both GONG and HMI could be associated with an EUV-bright region at approximately the same location. However, the converse was not true: some regions bright in EUV were not detected in the seismic maps. The results in Papers I and II were based on a qualitative comparison of EUV and seismic maps only, so no quantitative relationship between EUV brightness and seismic detectability could be established.

In this paper, we present results utilizing quantitative comparisons of the intensity of EUV emissions with the seismic detections from the HMI Doppler data for the same nine months of data analyzed in Papers I and II. Two distinct approaches are used. The first follows from the results in Papers I and II, which showed that of 27 regions that were large National Oceanic and Atmospheric Administration (NOAA) active regions (ARs) after appearing on the east limb (as seen from Earth), only about 50% were detected seismically on the far-side as they approached the east limb. In the work presented here, we calculate the integrated EUV intensities of these regions when on the far side to attempt to explain these results by determining whether there is a systematic trend in the EUV intensity of the regions detected seismically compared to those not detected. Far-side seismic detections determined by two separate processing techniques using the same HMI Doppler data are analyzed: the 2013 processing also analyzed qualitatively in Paper II and an improved 2014 processing that produces the seismic maps found at http://jsoc.stanford.edu/data/farside/ (Lindsey and Braun, 2017). The processing of the HMI Doppler data used to create these newer far-side seismic maps differs substantially from the processing used to produce the maps that were analyzed in Paper II (see Section 2.1). Overall, we find that, while there is a range of EUV intensities where regions may or may not be detected seismically using the HMI Doppler data, there appears to be an intensity level above which they are almost always detected and an intensity level below which they are never detected.

In the second approach, we analyze concurrent EUV and seismic far-side maps, using the improved seismic maps from the 2014 five-day processing only, for the same nine-month period. We find that 95% (21 out of 22) of the far-side seismic regions corresponds to an extreme ultraviolet plage that either becomes a NOAA-designated active or plage region when it reaches the east limb or was one before it crossed to the far-side hemisphere. The one region that was not associated with a designated NOAA region did correspond to a EUV plage with a low EUV intensity. For the set of seismically detected far-side regions analyzed here, we find a low but significant correlation between the area-integrated strength of the seismic signature and the area-integrated EUV intensity. A correlation was expected because previous research had shown that seismic signal strength correlates with magnetic field strength (González Hernández, Hill, and Lindsey, 2007; MacDonald *et al.,* 2015) and EUV had been shown to correlate well with photospheric magnetic fields *(e.g.*, Parker, Ulrich, Pap, 1998; Henney *et al.*, 2015). This work, however, to our best knowledge, is one of the first studies to compare quantitatively the seismic signal strength and EUV plage intensity of a far-side active region.

The outline of the paper is as follows. In Section 2, we describe the (helio)seismic and EUV data sets used. In Section 3, the techniques for identifying and computing the total intensity of bright plage regions and associating them with both NOAA active regions and seismic far-side regions are explained. In Sections 4 and Section 5, respectively, we present the results of the first and second approaches described above. Section 6 contains a summary and discussion of the results and implications for developing more accurate magnetic field Carrington maps for use in space weather prediction.

## 2. Seismic and EUV Data Sets

### 2.1. Seismic Carrington Maps and Far-side Strong Field Regions

Helioseismic holography reconstructs the acoustic field inside the Sun using the seismic disturbances observed at the surface and captured in the Doppler data. Active regions introduce a perturbation in the acoustic-wave travel time for waves reflecting off the far side (Lindsey and Braun, 2000; Braun and Lindsay, 2001) that can be measured by correlating acoustic signals of waves that impinge on far-side active regions with their returning echoes. The transit time for waves reflecting from magnetic regions is several seconds less than for waves reflecting from the quiet Sun. This induces a negative phase shift in the correlation on the order of tenths of a radian. Braun and Lindsey (2000) attributed this reduced travel time primarily to a reduced path length caused by a Wilson-like depression resulting from the plage magnetic fields. The far-side Doppler data is extremely noisy and, to increase the signal to noise ratio in the Doppler data, from 24 to 120 hours of Doppler data are used to create each far-side seismic map. Improving and diversifying seismic sensing of solar activity in the far hemisphere is an active area of research (Zhao, 2007; Ilonidis, Zhao and Hartlep, 2009).

Carrington maps showing regions of strong seismic signature are produced from HMI Doppler data on a 12-h cadence (at 0 and 12UT). For the results in Section 4, seismic Carrington maps created by two different processing pipelines are analyzed: the HMI seismic prediction maps created in 2013 analyzed in Paper II and newer seismic maps available through the Joint Science Operations Center (JSOC) at Stanford (http://jsoc.stanford.edu/data/farside/). One major difference is that the newer maps utilize five days of HMI data to create one seismic Carrington map; we will refer to this as the 2014 five-day processing hereafter. This processing results in far-side seismic (FS) regions that can be identified and followed for many days. The earlier 2013 processing utilized approximately two days of Doppler data, which frequently resulted in intermittent detections of strong field regions (*cf.* Paper II).

The top panel in Figure 1 shows a composite seismic Carrington map (0 to 360° in longitude, -90° to +90° in latitude) for January 18, 2012 at 12UT produced using the 2014 five-day processing. The far side portion of the map shows the seismic signature ($\tau$ is the travel-time shift in seconds relative to quiet Sun regions) and the Earth-side region shows the HMI magnetogram (with the magnetic field $B$ in gauss). The seismic signature used in the helioseismic analysis is the phase shift $\Delta$ determined from the helioseismic holography correlation analysis described above. The travel time $\tau$ (in seconds) shown in the figure is related to the phase shift $\Delta$ using a period of 5 minutes, *i.e.* $\Delta = 2\pi\tau/300$ radians.

The Carrington map in the lower panel of Figure 1 shows the far-side active regions detected seismically**.** Four far-side strong field regions are shown that were identified from the seismic signatures in the composite map in the top panel; each of these regions was tracked for several days. These far-side (FS) regions are labeled by year and number. The seismic signature strength $S$ of an FS region is the integration of the phase shift $\Delta$ over the area (in units of millionths of a hemisphere – $\mu$hs) where the phase shift exceeds 0.085 radian (see Lindsey and Braun, 2017). To be classified as a far-side seismic (FS) region, the seismic signature strength S must exceed the threshold of Stanford's Strong-Active-Region Discriminator, $S > 400$ in units of a millionths hemisphere radian ($\mu$hs rad). A text file is also created for each of the twice-daily maps listing the identified FS regions, their centroid locations on that map and their seismic signature strength $S$ **(**see http://jsoc.stanford.edu/data/farside/). These data files were used to obtain the results in Section 5 where far-side seismic regions were compared with concurrent far-side EUV plages.

## 2.2.  EUV Carrington Maps

We use STEREO EUVI observations in the 304Å passband to calculate the brightness of active regions on the far side of the Sun, using intensity of EUV 304Å as a proxy for magnetic field strength. To facilitate comparison with the seismic Carrington maps, we produce "all Sun" EUV Carrington maps on the same 12-h cadence (0 and 12UT) as the seismic maps by combining EUVI 304Å data from the two STEREO spacecraft with the data from the *Atmospheric Imaging Assembly* (AIA) on SDO in a similar 304Å passband. The procedure we used is based on the IDL procedure in the SolarSoft tree for STEREO ($SSW/stereo/ssc/idl/beacon/ssc_euvi_synoptic.pro), written by Wm. Thompson. It uses the STEREO/EUVI and SDO/AIA 304Å images closest in time to 0 or 12 UT; the STEREO/EUVI and SDO/AIA images, covering 360° in solar longitude, were generally taken within minutes of each other. We choose to use the 304Å passband because this chromospheric wavelength shows less variability with solar activity on the time scale of minutes and hours compared to the variability in the other EUV channels whose emission is from higher in the corona. These maps, in Flexible Image Transport System (FITS) format, are available at https://solarmuse.jpl.nasa.gov/data/euvisdo_maps_carrington_12hr/304fits/. The intensity data in these FITS files is in EUVI DN/s; the AIA data is converted to the same units *via* a calibration. In contrast to the exponentially scaled Carrington maps used for the qualitative results in Papers I and II, these new maps retain the full dynamic range of the original EUV data. This study covers the same two time periods used in Papers I and II: 2011 February through June, and 2012 January through April, when major active regions were reported by NOAA.

## 3. Methodology for Analysis of the EUV Data

### 3.1. Determining the Intensities of Plage Regions

An automated routine is used to find active regions based on the EUV 304Å intensity of the plages, where the selections are made independently for each individual twice-daily EUV Carrington map. For consistency over a series of maps, we scale the pixel intensity $C_i$ (in DN/s), to the median intensity of each map $C_m$, the latter reflecting the intensity of the quiescent Sun. The top panels of Figure 2 show this median intensity $C_m$ for both time periods. It can be seen that for the two periods in 2011 and 2012, the quiescent intensity is quite stable with a time average of 367±6 and 372±9 DN/s respectively. The bottom panels of Figure 2 show the histogram of the pixel intensity (normalized to the median of each map) of all the individual maps. Based on these plots, empirically, we consider a chromospheric plage region to have an intensity of at least three times the intensity of the quiescent Sun (see Qiu *et al.*, 2010, Figure 3b). Therefore, our automated routine for finding plages selects pixels whose intensity is greater than three times the median intensity $C_m$ of the entire map, and groups these pixels into spatially distinct plage regions, each enclosed in a rectangular box, using a multi-step procedure described in Appendix A. This automatic routine yields a selection of about one dozen plage regions on a Carrington map in a given day in 2011 or 2012. Figure 3 shows two sample Carrington maps with identified plage regions enclosed in white boxes.

For each Carrington map, we then calculate the properties of each boxed plage, including their positions, effective area and integrated intensity using the original 3601x1801-pixel map. The centroid of each region is calculated by $x_0 = \Sigma x_i (C_i - C_m) / \Sigma (C_i - C_m)$ and $y_0 = \Sigma y_i (C_i - C_m) / \Sigma (C_i - C_m)$, where $x_i, y_i$ and $C_i$ are the longitude, latitude and intensity (in DN/s) of a given pixel, and the sum is over all pixels within the rectangular box enclosing the plage. Here $x_i$ and $y_i$ are in radians. The effective area $A$ of the plage is the total of the spherical surface area of these "above threshold" pixels, normalized to $R^2$ where $R$ is the radius of the Sun. A plage determined in this way has an area of 0.01 to 0.20 steradian, which is equivalent to 800 to 16000 millionths of the total solar surface area. We normalize the plage areas to the area $A_0$ of a typical active region, which occupies about 15 degrees in longitude and 5 degrees in latitude, yielding $A_0 = 0.02$ steradians (see the review by van Driel-Gesztelyi and Green, 2015). Finally, the integrated intensity $I$ is the sum of the intensities of the ``above threshold'' pixels, normalized to the quiescent Sun intensity $C_m$, over the spherical surface area as

$$I = \sum \left[ \frac{C_i}{C_m} \frac{\cos (y_i) \Delta y_i \Delta x_i}{A_0} \right]. \qquad (1)$$

Our plage intensities calculated in this way are thus normalized to the intensity $C_m A_0$. With this normalization, the integrated plage intensity $I$ measured in this study ranges between 3 and 30 times $C_m A_0$, the integrated quiescent Sun intensity in an area of 0.02 steradians. It should be noted that throughout this paper, when we refer to the integrated EUV intensity of a plage, it always refers to integration over the plage area normalized to $C_m A_0$.

The top panel in Figure 3 shows a sample EUV map on 2011 February 14 at 12 UT, with ten numbered white boxes each enclosing a plage selected using the automatic procedure described in detail in Appendix A. They are sorted and numbered by centroid longitude. Since the plages and enclosing boxes are selected independently for each map, as the plages evolve by emerging, growing, or decaying, the size, total number and numerical sequencing of these boxes change, and the integrated intensity $I$ and centroid of the evolving plage also vary. Such variations are seen in the bottom panel of the figure, showing the plage regions on a Carrington map half a day later on 2011 February 15 at 0 UT.

### 3.2.  Association of Plage Regions with NOAA Active Regions

Paper II addressed the issue of the reliability of helioseismology for predicting large active regions prior to their appearance on the east limb of the solar disk (as seen from Earth). NOAA-designated active regions (ARs) are a subset of magnetic active regions that have been classified by NOAA personnel from the solar images of the Earth-side hemisphere as they appear on the east limb and are considered to be the most important magnetic field regions. The regions are listed in Solar Region Summary (SRS) files, available at [ftp://ftp.swpc.noaa.gov/pub/warehouse/](ftp://ftp.swpc.noaa.gov/pub/warehouse/). To analyze the ability of helioseismology to predict large NOAA active regions before appearing Earth side, we first searched the SRS files for the largest NOAA ARs with east longitude > 52° (that is, 38° from the east limb): February-June 2011 and January-April 2012 (see Paper II). To limit the number of ARs to analyze, we chose only ARs with sunspot area $A$ >82 (in units of millionths of the solar disk area) because this captured all of the active regions detected seismically as well as some others. We searched as far from the limb as east 52° because often a region is not designated by NOAA until it has rotated fully onto the near-side hemisphere and is less foreshortened. We had found 13 such large east-limb ARs (termed LEARs) in the 2011 data set and 14 in the 2012 data set. Of this set of 27 LEARs, it was found that only about 50% were detected at least once within the seven days prior to their appearance Earth side using the 2013 processing and the NSO/GONG seismic-signature discriminator. Many were predicted on only a few maps (Paper II). In this paper, we make a quantitative comparison of the EUV intensity of these same 27 LEARs when on the far side to attempt to better understand why only about 50% were detected seismically. In addition, we compare the detections made using the new five-day 2014 processing and the Stanford seismic-signature discriminator.

For an initial identification of a NOAA AR in our set of 27 LEARs with one of our EUV plages, we compare the centroids of the EUV plages (marked with "X" on the maps) with those of the NOAA ARs on the dates of their NOAA designations (marked with "o" on the maps). We note that, in a large active region, the separation between the positive and negative flux centroids is between 10 - 20 heliographic degrees (Wang and Sheeley, 1989; van Driel-Gesztelyi and Green, 2015). Therefore, we consider that if the difference in the position of an EUV plage from that of the AR satisfies $-15° \leq \Delta x \leq 15°$ and $-5° \leq \Delta y \leq 5°$, the plage is associated with the AR. The two solid white boxes in the top panel of Figure 3 show such identifications of ARs 11163 and 11164 on the EUV map from 2011 February 14 at 12 UT. This identification process was done for each map in the two time periods under study. Since we use EUV maps to identify active regions on the far side, the association is made only for 1 - 14 days before the AR appears Earth side.

Through visual inspection, we find that the success rate of the automatic AR-plage identification is about 70%. When different active regions are very close to each other, occasionally a plage is identified with the wrong AR; these misidentifications are corrected manually. More often, as nearby active regions grow and start to merge, they cannot be distinguished as different plages and the automatic routine then finds only one large plage. Figure 3 shows such a case: two distinct plage regions (boxes ``3'' and ``4'' on the top panel) are found on February 14. On February 15, they have merged and are recognized by the automated routine as only one region (box ``4'' on the bottom panel). This also happens when active regions cross the seam where the separate EUVI images from STEREO A and STEREO B are joined. This seam is visible in the maps in Figure 3. This is only a problem in the 2011 data set when the separation of the two STEREO spacecraft had just exceeded 180° and the center of the far-side hemisphere was not well observed, leading to the missing data (black regions near the southern pole) and stretching of the features seen at the seam, due to foreshortening, seen in Figure 3. In these situations, the routine uses the coordinates of a plage box enclosing this plage in an earlier or later map (the closest in time), when two plage regions can be distinguished, to calculate the properties of the plage region associated with the AR. We term these boxes the "corrected" boxes to distinguish them from the automatically determined boxes. Two supplementary movies, made from maps such as in Figure 3, are provided in the supplementary online material, 2011euv_lear.mp4 and 2012euv_lear.mp4. The movies show the EUV plage identifications and associations with the NOAA active regions for all the EUV maps during the two periods. The detailed explanation of the various labels and box types in the movies are given in Appendices A and B.

The calculation of evolving plage intensities using the dynamically determined boxes, corrected manually as necessary, is the primary method used in this paper. To estimate uncertainties and test the robustness of this primary method, we also measure plage properties in several different ways. In one approach, we measure the integrated intensity of a plage in a constant, rather than a time-dependent, plage box, and this constant box is chosen to be the largest box out of all dynamically determined boxes that enclosed the plage during the 14 days it was on the far side. This approach is discussed further in the next paragraph. In another approach, we integrate the difference intensity ($C_i$ - $C_m$) in all pixels in the box, not just the ``above-threshold'' pixels as we did in Equation 1. The details of these procedures and the majority of the comparison between these measurements and the primary method are presented in Appendix B.

Figures 4 and 5, for 2011 and 2012 respectively, show the time history of integrated EUV intensities of several plages associated with NOAA ARs for the two weeks before their appearance Earth side. In the plots, intensity is calculated by integrating intensities of ``above-threshold'' pixels using Equation 1 using three different methods to determine the plage box (*cf.* Section 3.1). In each plot, dotted curves show evolution of the integrated intensity $I$ from the automatic identification. Solid lines show the intensity integrated in the plage boxes corrected for misidentification or merging effects. Misidentification occurs sometimes for some ARs, for example, NOAA 11163 in Figure 4a, NOAA 11195 in Figure 4e, and NOAA 11402 in Figure 5a. These lead to large swings seen in these intensities in dotted lines (automatic) compared with the intensities calculated using the corrected boxes (solid lines). If no misidentification occurs, the solid curve and dotted curve are identical, *i.e.*

no correction is made. Finally, the dashed lines show the intensity measured in the same, constant box for a given AR, as described above. It is seen that the integrated intensity within the constant box is either the same or only slightly larger than that calculated in the time varying boxes which are dynamically identified. Measurements with yet other different methods, discussed in Appendix B, do not differ significantly from each other. Therefore, in this study, we only use the EUV intensities (normalized to $C_m A_0$) calculated using the primary method: intensities integrated in ``above-threshold'' pixels from boxed plage regions which are dynamically determined and corrected for misidentification or merging effects, i.e., the intensities plotted in solid lines in the figures.

### 3.3. Association of Seismic Regions with Plages

As described in Section 2.2, the 2014 five-day processing of the HMI magnetograms produces far-side seismic regions that can be tracked continuously for several days or more as they move across the far side (*cf.* Figure 1). Recall that the FS regions are designated by year and number. The data files contain the Carrington longitude and latitude of the centroid of the FS region and the seismic signature strength $S$ for each date and time it was identified. We use the same basic technique to associate EUV plages with seismic FS regions as we used to associate plages with NOAA ARs (Section 3.2). In each period of 2011 and 2012 that we analyzed, 11 far-side seismic regions were found and each region was detected for multiple days prior to its appearance on the Earth-side hemisphere. Our routine looks for the EUV plage at the same time whose centroid is closest to the centroid of the seismic region; an association is made if the centroids are within *±15°* in longitude and *±5°* in latitude of each other.

The top panel of Figure 6 shows the EUV map for 2012 January 18 at 12UT, the same date and time as the seismic maps in Figure 1. The contours enclosing each of the four far-side seismic regions, FS-2012-002 through -005 are overlaid on the EUV map. Solid boxes indicate that an association has been made between these plage boxes and the FS regions. FS-2012-003 and -004 have both been matched to the same box that encloses both. It can be seen that each of the seismic signatures is well within an EUV plage outlined by a rectangular box, suggesting that the automatic approach rather successfully identifies plages associated with seismic regions. Two additional movies, with one frame per map, are provided in the supplementary online material (2011euv_fs.mp4 and 2012euv_fs.mp4); these movies show all the associations of seismic far-side (FS) regions with EUV plages for the periods selected for analysis in 2011 and 2012. Additional information about the movies is given in Appendix C.

After this first automatic step, we make a visual inspection and make corrections for misidentified plage regions or for merging effects in the same way as was done for the EUV plage-AR association described in Section 3.2. That happens in only a few cases and is usually caused by an EUV plage becoming too large, either due to plages crossing the STEREO A-B seam or plages merging, causing the centroid of the plage to fall beyond our tolerance.

In some cases when two plage regions merge, however, we do not correct for the merging, but associate the merged large EUV region with the seismic region. This is because of the much lower spatial resolution of the seismic maps and resulting uncertainty as to whether one or both regions contribute to the seismic detection. The bottom map in Figure 6 (2011 February 20 at 12UT) shows the example of the 2011-FS-002 region associated with a large plage (solid white box 4), which encompasses two merging active regions (ARs 11163 and 11164). The contour of the FS region, taken from the FS region map for that day and time, is also shown in the figure. The temporal evolution of 2011-FS-002 shows that the seismic signature gradually moves from AR11163 to AR11164, hence both active regions contribute to this single seismic signature. By allowing the seismic regions to be identified with merging plage regions, the integrated EUV intensity becomes larger (Section 5) compared with the integrated EUV intensity identified with the two individual NOAA AR active regions (Section 4).

To test the robustness of our determination of the integrated EUV intensity of a plage associated with a particular FS region, we again calculated the integrated EUV intensity in several additional ways as described briefly in Section 3.2 and in more detail in Appendix C.

We found that the integrated EUV intensities were insensitive to the different methods. The results of the different determinations are discussed in Appendix C.

# 4. EUV Intensities of NOAA Active Regions and Comparison with Far-side Seismic Detections

In this section, we present the far-side integrated EUV intensities of the plages associated with our set of 27 large east limb NOAA active regions (LEARs) to search for systemic trends in the intensity of those regions detected seismically and those not detected seismically. Recall that these regions were selected on the basis of their large sunspot area ($A > 82$ in units of millionth of the solar disk area) on the date of their NOAA AR designation Earth side, not their magnetic flux or EUV intensity.

Table 1 presents the results of our analysis of the thirteen 2011 LEARs while on the far side, comparing the EUV intensities of the corresponding far-side region and indicating whether or not the region was detected seismically. The EUV intensities in the table are calculated using the primary method, described in Sections 3.1 and 3.2. Column 1 gives the NOAA AR designation; the NOAA designation for the previous Earth-side transit is given in parentheses, if applicable. Column 2 gives the date that the NOAA AR appeared Earth side. Columns 3 to 5 give the integrated intensity $I$ (Equation 1) and its standard deviation averaged over three different lengths of time. The third column gives the average over the two weeks prior to the corresponding NOAA AR appearing Earth side; the fourth and fifth columns gives the average over 1 to 7 and 8 to 14 days respectively. For abbreviation, a time-averaged EUV intensity is denoted as $\langle I \rangle$ hereafter. The intensities $I$ and $\langle I \rangle$ given in the text and in the tables are always in units of $C_m A_0$, the quiet Sun intensity of a region of size $A_0 = 0.02$ steradians (Section 3.1). The sixth column gives the number of days (2 maps per day) that the region was detected using the 2014 five-day processing of the HMI Doppler data and 0 days indicates the region was not detected. Note that four of the 13 LEAR regions in our set (31%) were detected seismically. In all cases when the region was detected, the detection was continuous for this number of days. The seventh column gives the number of days the region was detected using the 2013 processing of the same HMI Doppler data, which was the processing used in the analysis in Paper II; the detections were often intermittent. Here six of the 13 (46%) were detected on at least one map (0.5 days) during the two weeks prior to appearing on the Earth-side hemisphere.

The data in Table 1 is sorted by descending value of the time-averaged integrated EUV intensity $\langle I \rangle$ for the two-week period (column 3) prior to appearing Earth side. It can be seen that the first five regions listed had large mean EUV intensities ($\langle I \rangle > 8$ for the two-week average) before appearing Earth side and four of these were detected seismically using both data processing procedures. Intensity $I$ *versus* time curves for two of the four, ARs 11163 and 11176, were shown in Figures 4a and 4b, respectively. Note that the new five-day processing was able to track these four regions across the far side for over a week, with the apparent exception of the second one in the table, AR 11163, which was tracked for $\approx 3.5$ days. In fact, AR 11163 (NOAA coordinates Carrington longitude = 176°, latitude = 18°) was identified as FS-11-002 on 15 February and was being tracked when AR 11164 (NOAA coordinates Carrington longitude = 162°, latitude = 28°) emerged nearby on 14 February. Apparently, it grew to seismically dominate the older AR11163 over the course of 3 or 4 days: the centroid of the FS-002 detection moved slowly to the location of AR11164 over this time period. While the two plage regions can occasionally be separated in the EUV maps, only one region, FS-2011-02, was identified in the seismic maps. The bottom EUV map in Figure 6 has an overlay of the outline of FS-2011-02 from the seismic map for the same date and time (2011 February 20 at 12UT). It can be seen that the region of the FS signature encompasses the plages associated with both future NOAA regions, ARs 11163 and 11164.

The fourth region in Table 1, future AR11216 (remnant AR11195), was not detected using either processing. Its integrated intensity averaged over two weeks is quite comparable to those detected, but the seismic signature, while visible in some of the composite seismic maps, apparently never reached the strong active region threshold for either the 2014 five-day processing or the older 2013 processing. We also note that the light curve for AR11216, shown in Figure 4d, shows that AR 11216 decays, and its intensity averaged in the week before appearing Earth side has dropped to $\langle I \rangle = 5.8$, which is only a half of the average intensity a week earlier.

The last four regions in Table 1 all had very low mean two-week intensities ($\langle I \rangle < 2$) and were not detected by either processing procedure. All four also had $\langle I \rangle \approx 0$ for 8 to 14 days before appearing Earth side, indicating that they emerged during the 1- to 7- day period. The evolution of the intensity for one of these, AR 11226, was shown in Figure 4c. It clearly shows the emergence of the region only 1.5 days before appearing Earth side, followed by a rapid growth. From Table 1 alone, we can hypothesize that the four lowest average intensity regions of the 13 LEARs (31%), all of which emerged on the far side, were not detected seismically on the far side because they had not yet grown to be strong active regions before appearing Earth side.

The four regions in the middle of Table 1 (ARs 11228, 11227, 11183, 11195) have two-week average intensities in the range $2 < \langle I \rangle < 5$. Two regions, ARs 11228 and 11195 were detected using the older processing, but only for 0.5 and 1.5 days, respectively. AR 11195, which emerged five days before appearing Earth side, was detected with the older HMI processing for the last 1.5 days before appearing Earth side, when it was at its brightest ($\langle I \rangle \approx 8$). AR 11195's intensity *versus* time is shown in Figure 4e. AR 11228 was fairly bright ($\langle I \rangle \approx 10$) for the week before appearing Earth side (Figure 4f), but it was detected on only one seismic map (0.5 days) during this period. The five-day averaging of the 2014 processing might play a role in the failure to detect AR 11195. Or, the non-detections reflect the fact that the seismic maps are intrinsically much less sensitive than the EUV maps (Lindsey and Braun, 2017).

Table 2 presents the same data for the fourteen 2012 new large east limb NOAA active regions (LEARs) in our set, organized as in Table 1. Here again the regions are sorted by decreasing integrated EUV intensity averaged over the two weeks prior to the corresponding NOAA AR appearing Earth side. The first five high EUV-intensity regions in Table 2 (ARs 11410, 11423, 11395, 11408, and 11402) all had large EUV intensities ($\langle I \rangle \geq 8$) in the week before appearing Earth side and $\langle I \rangle > 6$ for the two-week mean) and all were detected seismically using both processing procedures. Among these regions, AR 11402 (Figure 5a) has the lowest average intensity ($\langle I \rangle$ =6.3 for the two week mean). It emerged 8 days before its appearance Earth side, and grew rapidly with a large average intensity in the last week ($\langle I \rangle$=10.4). The older 2013 processing detected this region for four days (Column 7).

The next four regions in Table 2 (ARs 11433, 11401, 11420 and 11471), with two-week average intensities in the range $3 < \langle I \rangle < 6$, were not detected in the 2014 five-day processing, but were detected seismically using the 2013 processing for 0.5 to 1.5 days. The intensity *versus* time curves for AR 11433 and 11471 are shown in Figure 5b and 5c, respectively. AR 11433 maintains a moderate intensity for two weeks ($I \approx 6$). For AR 11471, the intensity rises rapidly and is large ($I > 10$) for about 3 days before appearing Earth side, and the 2013 processing detections were during this period. Thus, the five-day averaging of the 2014 processing may have contributed to the region being missed in that processing, as suggested for the 2011 AR 11195 in Figure 4e.

The last five regions in the table (ARs 111429, 11445,11459,11432, AR 11434) all had very low mean intensities ($\langle I \rangle < 2$), all emerged on the far side, and all were not detected by either processing procedure. A small plage is visible at the location of the future AR 11434 about two days before appearing Earth side, but the plage is below our threshold. This non-detection of far-side regions with such low integrated intensities is consistent with the results seen in the 2011 data in Table 1. Here, as from Table 1, we can hypothesize that five of the 14 NOAA large east limb active regions (36%), all of which emerged on the far side, were not detected seismically before appearing Earth side because they had not yet grown into strong active regions.

From the results in Tables 1 and 2, we conclude that there is a correlation between far-side EUV plage intensity and helioseismic detectability of future NOAA ARs: the EUV plages with the largest mean intensities are generally detectable for several days to a week or more and those with the lowest mean intensities are not. However, we find a wide range in the middle where other factors are apparently at play in determining the helioseismic detectability. While both the seismic signature and the EUV plage intensity result from strong magnetic fields, the seismic signature responds to the photospheric magnetic fields whereas the EUV emission is from the chromosphere, and the physical processes causing the two signals are quite different. Also, the temporal scales of the data are different and the seismic maps have lower sensitivity than the EUV maps (Lindsey and Braun, 2017). The next section compares the strength of the seismic signatures to the EUV intensity for all seismic active region detections in the two periods under study, not just those associated with large future NOAA active regions (LEARs), to investigate more closely the relationship between the two signatures.

# 5. Comparison of Seismic Far-side Regions and EUV Plages

In this section, we consider the entire set of 22 far-side seismic regions found using the newer five-day 2014 processing that are within our 2011 and 2012 time periods.

Using the method described in Section 3.3, we have identified the EUV plage(s) associated with these 22 far-side regions and measured the integrated intensity $I$ of these EUV regions (normalized to the quiet Sun intensity integrated over an area of 0.02 steradian) for comparison with the FS seismic signature strength $S$ (in units of a $\mu$hs rad). The seismic signature strengths S for the FS regions were taken from the data files at http://jsoc.stanford.edu/data/farside/.  In addition, we have associated the FS-associated EUV plage region with a corresponding NOAA-designated magnetic region whenever possible, thus creating three-way associations.

## 5.1. Association of Far-side Seismic Signatures with EUV Plages and NOAA-designated Active Regions

Table 3 lists the 11 seismic far-side (FS) regions identified during the first period of data analyzed, February to June 2011. The first column is the designation of the FS region, the second column gives the number of days the region was detected seismically (2 maps/day), the third column gives the maximum (over the days of detection) seismic signature strength S of the FS and the mean seismic signature strength $\langle S \rangle$ averaged over the period of the detection with the standard deviation in parenthesis. Both $S$ and $\langle S \rangle$ are in units of $\mu$hs rad in the text and in the tables. The fourth (fifth) column gives the FS centroid Carrington longitude and latitude on the first (last) last date the region was identified.  Column 6 gives the mean integrated EUV intensity $\langle I \rangle$ of the plage associated with the FS region as described in Section 3.3; here the mean is calculated over the time period of the corresponding seismic detection and the EUV intensity is calculated using the primary method (*cf.* Section 3).

Column 7 of Table 3 gives the NOAA-designated regions that we associate with the FS-associated EUV plages. In some cases, it is a future NOAA active region (*e.g.*, a region that will become a NOAA-designated region after it crosses to the Earth-side hemisphere), and in other cases it is an old NOAA active region that has crossed to the far side hemisphere from the Earth side; we refer to the later as remnant NOAA active regions. Associating EUV plages with NOAA-designated regions is generally straightforward since the EUV plage region can be followed to (or from) the Earth side and compared with HMI magnetograms and other Earth-side observations.

Table 3 shows that every FS-associated EUV plage in 2011 can be further associated with an NOAA-designated region that will become a NOAA-designated active region when it reaches Earth side or was a NOAA-designated active region before it passed to the far side. Column 8 gives the coordinates of this associated NOAA-designated region, either on the date of its designation on the east limb or on the date it was last identified on the west limb before passing to the far side, as appropriate where the coordinates are taken from the SRS files. It can be seen that there is good agreement between the coordinates of the FS region and the associated NOAA-designated region.

Table 3, of course, includes the three seismic regions (FS-2011 regions 2, 4 and 12) associated with the future large east limb active regions (LEARs) discussed in Section 3. From the table, we observe that five additional regions (FS-2011 regions 3, 5, 7, 8 and 9) are also associated with future NOAA active regions. These NOAA-designated regions were not in our set of LEARs because they did not meet the selection criteria (sunspot area $A > 82$ when east longitude $> 52°$). The remaining three regions (FS-2011 regions 6, 10 and 11) are associated with regions that were remnants of NOAA-designated active regions that had crossed into the far hemisphere from the Earth side.

Table 4 contains the same information for the 11 far-side regions identified during the second period of data analyzed, January to April 2012, organized in the same way. The table includes five regions that are associated with future LEARs (FS-2011-53 and FS-2012-1, 2, 3 and 7), discussed in Section 3. Two far-side seismic regions (FS-2012-6 and 8) are associated with future NOAA-designated active regions that did not meet the LEAR criteria. One region (FS-2012-09) is associated with a future NOAA-designated plage. Figure 6 shows the contours of four of the 2012 FS signatures overlaid on the EUV map for 2012 January 18 at 12 UT; the seismic maps themselves were shown in Figure 1. Note the remarkably close correspondence between the seismic signature and EUV plage.

Here, in the 2012 data set, there is one region, FS-2012-10, that was not associated with an NOAA-designated active region or plage. It can be associated with a very weak EUV plage ($\langle I \rangle \approx 3$) which can be associated with a very small and short-lived bipolar magnetic region seen when the plage reaches Earth side. We could not associate this magnetic region with any past or future NOAA-designated region. The EUV plage for the days associated with FS-2012-10 can be seen in the accompanying movie 2012euv_fs.mp4 from April 8-10 at 0UT, after which it crosses to the Earth-side hemisphere and slowly fades away.

For the 2012 dataset, we also note that the two seismic regions FS-2012-004 and FS-2012-005 (visible in the lower map in Figure 6) are associated with the same EUV plage, which we identify as merged remnants of ARs 11388 and 11389. The smaller, shorter-lived FS-2012-004 starts just outside the plage (12 hours earlier than FS-2012-005 is detected), but moves into the plage in subsequent maps (see supplemental movie 2012euv_fs.mp4). This region was detected for four days. FS-2012-005 is located at the center of the plage box. This FS persisted for seven days, and its centroid also exhibited a significant shift during its detection. In the movie, two distinct EUV plages can be seen at the location of the seismic region a few days before the seismic detection, and, using Earth-side observations, these can be clearly identified as remnants of active regions NOAA 11388 and 11389. However, during the time of seismic detection, these two EUV plages have merged into a single large plage, and cannot be divided into two regions for comparison with the two FSs. Therefore, in the following section when we compare the strength of seismic signatures with the EUV intensities of associated plages, we combine the strength of these two seismic signatures, FS-2012-004 and FS-2012-005, both associated with the same merged plage.

From Tables 3 and 4, we find that all 22 far-side seismic regions analyzed can be associated with an EUV plage at essentially the same location. From the space weather prediction point of view, this fact implies that there were no false detections of far-side active regions in the time period analyzed when the new 2014 five-day processing was used, compared to about 10% false detections using the older processing, which produced intermittent predictions (*cf.* Papers I and II).

## 5.2. Comparison of Seismic Signature Strength and Plage EUV Intensity

In Section 5.1, we saw that each of the 22 far-side seismic regions (FSs) could be associated with a far-side EUV plage and that all but one of these (FS-2012-10) could be further associated with either a past or future NOAA-designated magnetic region. All but one EUV plage has a mean intensity $\langle I \rangle > 7$ during the detection of the seismic signatures, suggesting a correlation between plage intensity and seismic signal strength. Here, we compare the strengths of the EUV and seismic signatures directly.

Figure 7 shows several examples of the time evolution of the integrated EUV intensity of a plage associated with an FS region. To test the robustness, the integrated EUV intensity $I$, calculated using three different choices for the boxes, is shown in the figure (*cf.* Figures 4 and 5): the primary method (dynamic boxes visually corrected – solid line), the automatically identified boxes with no corrections (dotted), and the same fixed largest box (dashed) (see Sections 3.1 and 3.2). The first two kinds of light curves are plotted for the duration of the seismic region detection. The EUV light curve from the same, constant box is plotted for the entire 14 days before Earth-side appearance to show the EUV intensity prior to seismic detection as well. Measurements of EUV intensities from the dynamically determined corrected boxes (primary method) and from a constant box are not significantly different for the same day and time, indicating that the integrated EUV intensity of a plage is rather robust using our thresholding technique. Comparison with other methods of calculating the EUV intensity are discussed further in Appendix C.

Also shown in Figure 7 is the temporal evolution of the corresponding FS signature strength $S$ (thick solid blue line; scale on right axis). Comparing the temporal evolution of the seismic signature strength $S$ and EUV intensity $I$, it can be seen that in about half the cases (*e.g.*, Figures 7a-d), the seismic signature strength appears to evolve roughly along with the EUV intensity over a few days. However, the seismic signature strength drops faster than the EUV intensity when approaching the east limb. For the other half of the cases, such as those in Figures 7e and 7f, no such correlation in the time evolution is observed. As discussed at the end of Section 4, we attribute the lack of correlation to a variety of reasons relating to both differences in the physical processes producing the signatures and the differences in the data and data processing.

Using Tables 3 and 4, comparing the mean seismic signature strengths $\langle S \rangle$ (column 3) and mean EUV intensities $\langle I \rangle$ (column 6), we find a much greater variation in the seismic signature strengths than in the EUV intensity. The strength of seismic signature ranges between 100 and 1500, and the EUV intensities cluster between 7 and 30. The seismic strengths and the EUV intensities for both 2011(squares) and 2012 (triangles) are compared in Figure 8 in two ways. The mean EUV intensities $\langle I \rangle$ and seismic signature strengths $\langle S \rangle$, both averaged over the time of the seismic detection, are plotted in Figure 8a. Because the seismic signature strength $S$ is determined using five days of data, we also compare the mean EUV intensity $\langle I \rangle$ with the maximum (over the time of detection) seismic signature strength (Figure 8b). Despite the large scatter in the data, these plots indicate an overall trend that stronger seismic signatures are usually related to EUV plages of larger intensities.  The Spearman's rank correlation coefficients are 0.53 and 0.52 for the mean and maximum seismic signature strength, respectively, with the significance of its deviation from zero less than 3%. Thus, the correlation between the EUV and seismic signatures is significant.

In Figure 8a, the data seem to form two groups, as indicated by the two dashed ovals. For seismic signatures $\langle S \rangle > 500$, the mean EUV intensity $\langle I \rangle > 20$, and for $\langle S \rangle < 500$, $\langle I \rangle < 20$. For 2011, it appears the mean EUV intensity has a lower threshold at $\langle I \rangle = 10$, but for 2012, there are two EUV regions with lower intensity at $\langle I \rangle$ =7 (FS-003) and $\langle I \rangle$ = 3.3 (FS-010), respectively.

One puzzle in the 2012 data set (see Table 4) is FS-2012-10 that has a midrange mean seismic signature strength $<S>$ = 361, whereas the associated plage has a very low EUV intensity ($\langle I \rangle \approx 3.3$) and is the one plage not associated with a NOAA-designated region. (Recall from Section 5.1 that FS-2012-10 is associated with a small, short-lived magnetic bipole seen when the region reaches Earth side.) The EUV intensity of FS-2012-10 is larger before the seismic detection – and decays during the detection (see Figure 7f). Seismic region FS-2012-003 has a similar mean seismic signature strength (408) and modest EUV intensity ($\langle I \rangle$=7). It is associated with LEAR AR 11408, which is one of the large NOAA active regions and was detected seismically in the older processing as well (*cf.* Table 2 and discussion in Section 4). Here we have two far-side regions with similar seismic signature strengths and low or modest EUV intensity, yet one fails to become a significant magnetic region on the Earth-side, whereas the other becomes a major active region. These two exceptional cases, as well as the lack of a strong correlation between seismic and EUV signal strengths, again indicate that other factors enter into the relationship between magnetic fields, EUV emission, and seismic signatures as discussed at the end of Section 4.

The EUV plage intensities recorded in Tables 3 and 4 and plotted in Figure 8 are measured using the primary method. To test the robustness of our comparison, measurements of the EUV intensity calculated using several different methods (including those shown in Figure 7) are also compared with the seismic signature strength S. The EUV-seismic comparison using these different measurements does not exhibit significantly different trends than presented in Figure 8, as discussed in Appendix C.

## 6. Summary and Discussion

We have used STEREO far-side extreme ultraviolet observation of solar activity to study the reliability of helioseismic detection of far-side active regions. Two approaches were used to compare and analyze approximately six months of STEREO/EUVI and SDO/HMI data. Results of the two approaches are consistent and lead to the conclusions that i) the far-side seismic regions, as determined using the improved 2014 five-day processing, are always at the location of EUV plages when concurrent observations are compared, and ii) there is a low but significant correlation in the strengths of the two signatures with significant scatter. While both signatures result from strong magnetic fields, differences in the way these signals respond to the magnetic fields are apparent.

In the first approach used, we determined whether or not new large east-limb active regions were detected seismically on the far side before they appear Earth side and we studied how the ability to detect these regions seismically relates to the region's integrated extreme ultraviolet intensity on the far side. For the new five-day helioseismic processing of HMI Doppler data, we found that, while there is a range of EUV intensities for which far-side regions may or may not be detected seismically, there appears to be an intensity level above which they are almost always (≈90%) detected and an intensity level below which they are never detected. In Paper II, we reported that only about half of the new large east-limb active regions were detected seismically while on the far hemisphere. Based on the results in this work, we can attribute the lack of seismic detection of about a third of the these to the fact that the magnetic regions had apparently not grown large enough before crossing into the near hemisphere.

In the second approach, we analyzed concurrent extreme ultraviolet and helioseismic far-side observations for the same nine-month period and compared the strengths of the seismic and EUV signatures. We found that all 22 of the seismic far-side regions identified using the improved 2014 five-day processing could be associated with EUV plages at the same locations. Moreover, 95% (21 of 22) of these far-side seismic regions correspond to plage that either becomes a NOAA-designated magnetic region when it reaches the east limb or was one before it crossed from the Earth side to the far side. The far-side seismic region not associated with a NOAA-designated region could be associated with a low intensity plage and a small magnetic bipole. Thus, there were no false detections of far-side active regions among the set of 22 far-side seismic regions detected with this five-day processing, an important result for predicting space weather. A low but significant correlation with a broad scatter was found between the seismic signature strength and the EUV intensity of the associated far-side plage. The data also suggests a bimodal distribution of seismic signature strength and EUV intensities that is not understood. While both EUV and seismic signatures result from strong magnetic fields, many factors may influence the lack of strong correlation between the two: i) the EUV emission is from the chromosphere whereas the seismic signature relates to the photospheric fields; ii) the two signatures are created by different physical processes; iii) the spatial and temporal scale of the data are different; iv) the seismic maps are known to be inherently less sensitive (Lindsay and Braun, 2017); and v) there may be systematic effects in the seismic data processing. Clearly more work is needed to understand the relation between magnetic fields and seismic signatures.

Reliable, helioseismic predictions of far-side active regions are of great interest for space weather prediction because, in that case, only near-Earth observations are needed. Far-side detections help space weather predictions of future solar activity in detecting active regions before they appear Earth side where flares and CMEs have the largest effect on space weather at Earth. A second use for far-side detections of magnetic regions is to improve the synoptic magnetograms used as the boundary condition for heliospheric models such as Enlil (Odstrcil *et al*., 2004) which is routinely used for space weather predictions (http://www.swpc.noaa.gov/products/wsa-enlil-solar-wind-prediction). New active regions detected on the far side seismically could be assimilated into the current generation of adaptive synoptic magnetograms (Hickmann *et al.*, 2015; Upton and Hathaway, 2014a,b). These models already include flux transport from diffusion and differential rotation, but could be improved using the seismic detections of far-side active regions as demonstrated by Arge *et al.* (2013). A third use of far-side detections of active region strength is for the predictions of solar spectral irradiances. The unsigned magnetic flux is key input to the model predictions of the integrated solar spectral irradiance (Fontenla *et al.,* 2009) and F10.7 cm and EUV emission (Henney *et al.*, 2015) that are used for ionospheric and thermospheric modeling.

Ugarte-Urra *et al.* (2015) have measured the temporal evolution of EUV intensity in ten isolated active regions using all-Sun Carrington maps. They found that the duration and peak EUV intensity are correlated. They also determined a power-law relationship between an active region's integrated EUV intensity in the 304Å passband and the active region's magnetic flux. They demonstrate that, using this information, the evolution of the EUV intensity in plages can be used to predict magnetic flux in active regions yet to cross to the Earth-side hemisphere. A power-law relationship between the SDO/AIA 304Å intensities and the SDO/HMI magnetic flux was determined directly (*cf.* Figure 2 of Ugarte-Urra *et al.*, 2015); a calibration between STEREO/EUVI and SDO/AIA 304Å intensities then allowed them to determine indirectly the relationship between intensity in EUVI 304Å and magnetic flux (Ugarte-Urra, private communication). The correlations they found between the SDO/AIA 304Å intensities and SDO/HMI magnetic fluxes were much higher than the correlations we found here between STEREO/EUVI 304Å intensities and seismic signature strengths. Comparing the logarithm of the EUV intensity to the logarithm of the magnetic flux, Ugarte-Urra *et al.*, (2015) found a correlation coefficient >96%, whereas our comparable correlation coefficient was about 50%, as discussed (*cf.* Section 5.2).

In this paper, we have also measured the EUV intensity in 21 plages. The majority of active regions analyzed in this study are not isolated; AR11193 is the only active region studied in this paper and in Ugarte-Urra *et al.* (2015). Figure 9 shows the evolution of the integrated EUV intensity in automatically and dynamically determined boxes (*i.e.* boxes not manually corrected) for this active region over a few months, extending beyond the period of this study. Ugarte-Urra *et al.* (2015) have measured the EUV intensity as integrated counts in a fixed rectangular region of 20 by 20 degrees, which is quite a different approach from ours, as described in Section 3, which uses dynamically evolving boxes. Nevertheless, the evolution of the EUV intensity by our measurements is consistent with Ugarte-Urra *et al.* (2015), with the same duration of 40 days, and the same rapid rise and slower decay at similar times. Our present semi-automatic method, described in detail in Appendix A, can be further improved to identify and track individual active regions, and to examine whether all active regions exhibit the universal shape of evolution as reported by Ugarte-Urra *et al.* (2015). Such a capability would certainly aid the effort of full-Sun modeling and space weather forecasting.

**Acknowledgments** We would like to thank Jeffrey R. Hall and W. T. Thompson for processing the EUV data and creating the Carrington Maps. The work of PCL was conducted at the Jet Propulsion Laboratory, California Institute of Technology under a contract from NASA. The work of JQ was supported by the faculty sabbatical program of Montana State University. The work of CL was supported by grants from the Small Business Innovation Research Program of the NOAA and a contract with the Sun-Earth Connection Guest Investigator Program of NASA. The STEREO/SECCHI data used here are produced by an international consortium of the Naval Research Laboratory (USA), Lockheed Martin Solar and Astrophysics Lab (USA), NASA Goddard Space Flight Center (USA) Rutherford Appleton Laboratory (UK), University of Birmingham (UK), Max-Planck-Institut fur Sonnensystemforschung (Germany), Centre Spatiale de Liege (Belgium), Institut d'Optique Theorique et Applique (France), Institut d'Astrophysique Spatiale (France). The HMI data were provided by NASA through the Joint Science Operations Center for the SDO project at Stanford University.

**Disclosure of Potential Conflicts of Interest** The authors declare that they have no conflicts of interest.

**Appendix A: Method for Selection and Characterization of EUV Plage Regions**

We have developed a multi-step, semi-automated procedure to select EUV plage regions and track their evolution. During its lifetime, a plage is observed to grow and then decay in its intensity (in units DN/s per pixel in an EUV FITS map) as well as its area, similar to the evolution of magnetic field in an active region (Ugarte-Urra *et al.* 2015). Often two or more plages are close to each other. For these reasons, our method uses an automated thresholding to identify and track individual plage regions as they evolve, and to distinguish different plage regions close to each other. We consider that the intensity $C_i$ (in DN/s*)* of a plage pixel should be more than three times the median $C_m$ of the map (*e.g.* Qiu *et al.*, 2010), the latter representing the quiet Sun intensity. The plage threshold is therefore 3 $C_m$ in this study (*cf.* Figure 2 and discussion).

In the first step, we first identify the core of each separate plage regions: for each EUV map of 3601 by 1801 pixels evenly spaced in longitude $x$ [0 to 360°] and latitude $y$ [-90° to +90°] (Figure A1a), we make a binary mask: all pixels of intensity larger than the threshold 3 $C_m$ are given the value one and the rest are set to zero (Figure A1b). The binary map is then smoothed with an $N \times N$ binning. Regions inside the contour level $\alpha$ of the smoothed map are selected as individual plages. Through many tests of various $N$ and $\alpha$ values, we find that the combination of $N = 40$ and $\alpha = 0.5$ gives an optimal result per our visual inspection of the plage evolution (see the supplemental 304Å movies). Most of the time, this combination groups EUV bright pixels which are clustered together into one plage region, and, at the same time, allows separate distinct clusters of bright pixels to be grouped into separate distinct plage regions rather than into one much larger plage. Figure A1c shows the identified plage regions as a result of this procedure: each contour (of level 0.5) in the figure outlines the core of a separate plage region.

We next need to define a plage boundary to separate nearby plages and to enable us to determine the intensity of a plage by summing the above-threshold pixels. Some nearby above-threshold pixels will lie outside the contour because it was created on the $N \times N$ smoothed map, so we want to extend the boundary somewhat to include some of these bright pixels as well. A simple way to define the boundary is to use a rectangular box that is determined by the *x* and *y* ranges of the 0.5-level contour; we then expand the box by 20% in both *x* and *y* dimensions. The outer boundaries of each plage region determined in this way are illustrated by the rectangular white boxes in Figure A1c and d.

Finally, to remove sporadic local brightenings of small features, we also require that a plage should have a size greater than 20 × 20 pixels on the original 3601 × 1801 FITS map. This automatic plage selection method yields selection of about one dozen ``boxed'' plage regions on a Carrington map in a given day in 2011 or 2012. Figure A1d shows these boxes enclosing plages plotted on the original map, superimposed with contours of 3 times the median intensity, which is used as the threshold for plage pixels. Because of the smoothing and enlargement procedures, each plage box extends a little beyond the threshold contour of three times the median intensity $C_m$. This figure also shows clearly the necessity of the smoothing step as well as the boundary expansion.

Section 3.2 described our method for associating our set of 27 NOAA large east limb active regions (LEARs) with plage regions, identified as above. Two movies, made from maps such as in Figure 3, are provided in the supplementary online material, 2011euv_lear.mp4 and 2012euv_lear.mp4. The movies show the time sequence of the STEREO-AIA full-Sun maps for the two periods in this study with a time cadence of 12 hours. Each map is normalized to its median and displayed with a logarithmic scale for clarity. The movies show the evolution of the plages over the months analyzed and also the association between ARs and plages. In each map, a white box denotes the outer boundary of an automatically identified plage region, and the ``x'' symbol inside the box marks the centroid of the plage region. Solid white boxes indicate plage regions that have been associated with an AR in our set. Thus, the same plage region can be identified and tracked (with a white dashed box) for a few weeks to a few months, and when it is associated with one of the 27 LEARs during the two-week period before that active region appears Earth side, the plage region is outlined by a solid white box instead of a dashed white box. In the movies, the location of each LEAR is marked with the symbol ``o'' for 14 days before its Earth-side appearance, and the label (the AR number) is placed at the AR's longitude.

As briefly explained in Section 3.2, the automatic plage-NOAA identification sometimes fails when a plage region grows very big so that its centroid position is different from the NOAA position by more than our designated tolerance (15° in longitude and 5° in latitude). Such errors can be picked out and corrected simply through visual inspection by experienced researchers. Sometimes, when two originally separate yet close-by plage regions grow or when they cross the seam, they appear to merge and to be identified by the automatic approach as one large region. Again, all these cases are picked out manually through visual inspection of the movies. In these situations, the correction is not straightforward. Note that the automated routine identifies plage regions for each EUV map separately. During the rise and decay of close-by plage regions, we find some nearby plage regions that at some point in time are automatically identified together as a large merged region, whereas earlier or later they are automatically identified as separate regions. Therefore, our routine selects a time either before or after, whichever is closer to the time when the regions are merged, and uses plage boxes from the earlier or later time to replace the automatically identified "merged" plage box at the time of merging. In the supplemental movies, the replacement at a given time is seen as a black box. In this way, we can minimize uncertainties in measuring the plage properties caused by the merging or dispersing.

**Appendix B. Comparison of Different Calculations of EUV Plage Intensity**

The primary method for measuring plage properties used in this study was described in Section 3 and, in more detail, in Appendix A. Using the expressions in Equation 1 and the text in Section 3.1, we measured the integrated intensity $I$, the centroid, and the effective area $A$ for each plage region associated with an AR using the solid white or solid black box for each time during the 14 days before the AR's Earth-side appearance. These measurements of the integrated EUV intensity determined by the primary method are used throughout the study and are shown in solid curves in Figures 4, 5, and 7, and in the plots in Figure 8. To test the robustness of this method, we have calculated the integrated plage intensities by several alternative methods and compared them with those from the primary method.

First, we evaluate the sensitivity of the measurements to the outer boundary (outlined by the box) of the plage regions. Here, we calculate the integrated intensity $I$ using Equation 1, still summing only over pixels above the $3C_m$ threshold, but using constant, rather than dynamically varying boxes over the 14 days before the AR's Earth-side appearance. For this test, we chose the constant box in three additional ways: the largest among all dynamically determined boxes over the 14 days (called the ``same-big-box" hereafter), a large box that encloses all dynamically determined boxes over the 14 days (``include-all-box"), or the smallest box (the ``same-small-box") among all dynamically determined boxes. In the movies, a green box around a plage region denotes the ``same-big-box" for this plage region, and the other two kinds of constant boxes are not shown. Results using the same big box were also shown in Figures 4 and 5 (labeled "same box").

The measured integrated intensity for these three different choices of boxes, in comparison with the measurements from dynamically determined boxes (uncorrected and corrected, *cf.* Section 3) are shown in the top panels of Figure A2 and A3 for the time series of plage boxes associated with NOAA active regions 11423 and 11183, respectively. Comparing these curves, we find that integrated EUV intensity $I$ measured from the two types of constant large boxes (green and violet curves) does not vary significantly from the primary method - dynamically determined and corrected boxes (red curves). For the small box (yellow curves), $I$ does not vary if the plage does not grow or decay in area substantially during the two weeks on the far side, such as in the case of NOAA 11423 (Figure A2). However, when the area of the active region grows or decays significantly, the small box fails to capture such a rapid evolution and the intensity exhibits a large difference from other methods. The case of NOAA 11183 (Figure A3) is such an example of a growing active region.

Next, we investigate the effect on the intensity calculation of an alternative method that does not use a threshold intensity. In a previous study (Ugarte-Urra *et al.* 2015), the EUV intensity, with the quiescent intensity subtracted, is summed over all pixels in an a priori fixed region of an EUV Carrington map. To compare with this study, we also measure EUV intensities in each plage by

$$I' = \sum \left[ \frac{(C_i - C_{\mathrm{m}})}{C_{\mathrm{m}}} \frac{\cos (y_i) \Delta y_i \Delta x_i}{A_0} \right], \qquad (2)$$

where $C_i$, $x_i$, and $y_i$ are the pixel intensity, longitude, and latitude of the $i$th pixel, $C_{\mathrm{m}}$ is the median intensity of each map, and $A_0 = 0.02$ steradian is the same normalization constant as in Equation 1. In this method, the sum is over all pixels in the box, not just pixels exceeding the threshold. Further, we measure $I'$ in several different ways using the various types of boxes used above: dynamically changing boxes (uncorrected or corrected), ``same-big-box'', ``same-small-box'', and ``include-all-box''. The bottom panels of Figures A2 and A3 show examples of these measurements of intensity. The comparison of $I'$ measured in different boxes is very similar to that of $I$ using the thresholding method.

We also compare measurements from the two methods using (Figures A2 and A3, top panels) or not using (bottom panels) the threshold. We find insignificant differences between the $I$ and $I'$ measured using the same box. But when the box becomes larger, $I'$ increases faster than $I$, as more pixels of weaker emission are included, whereas the number of above-threshold pixels does not grow much. In this regard, the method of integrating the EUV intensity of the above-threshold pixels is less sensitive to the size of the box, and is therefore more robust.

More importantly, thresholding is critical to help distinguish nearby and evolving plage regions by confining the outer boundary of a plage region using the semi-automated and relatively objective approach; no *a priori* assumption about the region is necessary. The integrated EUV intensity $I$, calculated using the threshold in dynamically varying boxes captures evolution (including growth and decay) of an active region; therefore, this intensity is used throughout the paper.

## Appendix C. Comparison of Plage Intensity with FS Signature Strength

These different ways to measure the EUV intensity of plage regions associated with active regions are also used to measure the intensity of plages associated with far-side seismic FS regions (cf. Sections 3.3). The two movies 2011euv_fs.mp4 and 2012euv_fs.mp4 show the association of the EUV plages with the seismic signatures for every 12 hours during the two time periods of the study. In the movie, EUV plages are marked by white dashed boxes and their centroids marked by "x", as described in Appendix A. A plage associated with a seismic signature, whose centroid is marked by an "o" and label given at its longitude, is outlined by a white solid box. A black box indicates the correction of merging or seam effects, same as the corrections made to the NOAA association as described in Section 3.2 and Appendix A. The green box around a plage indicates a constant box, which is the largest among all boxes during the seismic detection, used to calculate the EUV plage intensity.

Figure 8 compared the mean intensities $\langle I \rangle$ of plages associated with FS regions to the mean seismic signature strengths $\langle S \rangle$; both signals were averaged over the time of the seismic detection. Figure A4 shows the same comparison, but now using eight different measurements of the mean EUV intensity of the plage, described above, versus the FS seismic signature strength $\langle S \rangle$. In each panel, black and red symbols show measurements of $\langle I \rangle$ and $\langle I' \rangle$, respectively, and measurements are made in dynamically determined boxes, a constant large box, a constant include-all box, and a constant small box (see previous section). Vertical and horizontal bars indicate the standard deviation of the data points during the seismic detection. It is seen that the trend of the EUV intensity versus seismic signature strength remains very similar.

For the 2011 and 2012 data taken together, the Spearman's rank correlation coefficients are 0.52, 0.52, 0.49, 0.58, respectively, for the $\langle S \rangle$ and $\langle I \rangle$ measured in varying boxes, constant big box, constant include-all box, and constant small box. respectively. For $\langle S \rangle$ and $\langle I' \rangle$ measured in the different boxes, the Spearman's rank correlation coefficients are 0.48, 0.47, 0.47, 0.57, respectively. Using the maximum seismic signature strength yields the correlation coefficients 0.50, 0.54, 0.52, 0.51 for $\langle I \rangle$, and 0.50, 0.53, 0.54, 0.49 for $\langle I' \rangle$. The significance of their deviation from zero is smaller than 3%. Therefore, the correlations are significant. Given the uncertainty in the measurements, we find all these measurements (except in the constant small box for the reason given in Appendix B) are equivalent in terms of their capability to calibrate the seismic signature strength or to predict the presence of active regions in the far-side hemisphere.

**Table 1. Seismic Detections of 2011 Large East Limb Active Regions**

| AR Number (AR Number previous rotation) | Date Visible Earth side | Mean EUV intensity[a] 1 – 14 days before Earth side | Mean EUV Intensity[a] 1-7 days before Earth side | Mean EUV Intensity[a] 8-14 days before Earth side | Days detected by HMI (2014 5-day processing) [FS designation] | Days detected by HMI (reprocessed 2013) |
|---|---|---|---|---|---|---|
| 11176 (11165) | 3/22/2011 | 22.4±2.9 | 22.6±3.8 | 22.2±1.8 | 11 [FS-011-004] | 10 |
| 11163 | 2/25/2011 | 17.2±4.4 | 17.9±3.9 | 16.5±4.9 | ~3.5 [FS-011-002[b]] | 2 |
| 11236 | 6/14/2011 | 15.4±6.7 | 15.9±2.8 | 15.0±9.2 | 9.5 [FS-011-012] | 10 |
| 11216 (11195) | 5/15/2011 | 8.9±4.6 | 5.8±3.9 | 12.1±2.8 | 0 | 0 |
| 11164 | 2/26/2011 | 8.3±6.2 | 13.3±4.2 | 3.4±3.1 | ~7 [FS-011-002[b]] | 6 |
| 11228 | 5/29/2011 | 5.2±4.8 | 9.4±2.3 | 1.0±1.9 | 0 | 0.5 |
| 11227 | 5/29/2011 | 3.6±3.1 | 4.3±2.2 | 2.8±3.6 | 0 | 0 |
| 11183 | 3/28/2011 | 2.8±3.1 | 5.3±2.4 | 0.2±0.9 | 0 | 0 |
| 11195 | 4/19/2011 | 2.1±3.1 | 4.5±3.1 | 0±0 | 0 | 1.5 |
| 11191 | 4/12/2011 | 1.4± 1.6 | 2.6±1.4 | 0.3±0.7 | 0 | 0 |
| 11166 | 3/3/2011 | 1.0± 1.8 | 2.0±2.1 | 0±0 | 0 | 0 |
| 11226 | 5/28/2011 | 0.4± 1.2 | 0.7±1.7 | 0±0 | 0 | 0 |
| 11193 | 4/13/2011 | 0.4±0.9 | 0.7±1.2 | 0±0 | 0 | 0 |

[a] Normalized to $C_m A_{0,}$ (see text)
[b] Two LEARs are associated with this FS region

**Table 2 – Seismic Detections of 2012 Large East Limb Active Regions**

| AR Number (AR Number previous rotation) | Date Visible Earth side | Mean EUV intensity[a] 1–14 days before Earth side | Mean EUV Intensity[a] 1-7 days before Earth side | Mean EUV Intensity[a] 8-14 days before Earth side | Days detected by HMI (2014 5-day processing) [FS designation] | Days detected by HMI (reprocessed 2013) |
|---|---|---|---|---|---|---|
| 11410 (11392) | 1/26/12 | 22.1±2.7 | 20.0±1.9 | 24.2±1.6 | 12.5 [FS-12-002] | 11.5 |
| 11423 (11410) | 2/22/12 | 15.2± 3.5 | 13.3±3.2 | 17.1±2.8 | 8 [FS-12-007] | 4.5 |
| 11395 (11376) | 1/6/12 | 14.8±5.3 | 14.8±5.3 | 0±0 | 6 [FS-11-053] | 1 |
| 11408 (11390) | 1/20/12 | 8.5±1.7 | 7.8±1.4 | 9.3±1.6 | 5.5 [ FS-12-003] | 4 |
| 11402 | 1/14/12 | 6.3±5.4 | 10.4±2.5 | 0.5±1.1 | 3.5 [ FS-12-001] | 4 |
| 11433 (11420) | 3/10/12 | 6.1±1.8 | 6.9±1.1 | 5.2±2.1 | 0 | 1 |
| 11401 | 1/14/12 | 5.4±3.3 | 7.7±1.7 | 2.2±2.1 | 0 | 1.5 |
| 11420 | 2/12/12 | 5.2±2.2 | 4.5±1.4 | 6.0±2.6 | 0 | 1 |
| 11471 | 4/27/12 | 3.1±4.0 | 6.2±3.5 | 0±0 | 0 | 1.5 |
| 11429 | 3/3/12 | 0.7±2.3 | 1.6±3.3 | 0±0 | 0 | 0 |
| 11445 | 3/23/12 | 0.6±2.5 | 1.3±3.4 | 0±0 | 0 | 0 |
| 11459 | 4/14/12 | 0.6±1.2 | 1.3±1.5 | 0±0 | 0 | 0 |
| 11432 | 3/9/12 | 0.5± 0.8 | 0.9±0.9 | 0.1±0.4 | 0 | 0 |
| 11434 | 3/10/12 | 0±0 | 0±0 | 0±0 | 0 | 0 |

[a] **Normalized to $C_m A_0$ (see text)**

**Table 3. 2011 Far-side Seismic Regions and NOAA AR Associations**

| FS in 2011 | days detected seismically | Seismic Signature Strength S [a] – max, mean (standard deviation) | FS Start Coordinates (Carrington Longitude, Latitude) | FS End Coordinates (Carrington Longitude, Latitude) | EUV intensity $I$ [b] over time of FS, mean (stdev) | NOAA AR association? | NOAA AR Coordinates (Carrington Longitude, Latitude) [c] |
|---|---|---|---|---|---|---|---|
| FS-2011-002 | 10.5 | 789, 509 (185) | 171°, 18° | 158°, 28° | 27.3 (8.8) | LEAR AR 11163 to 11164 [d] | (176°,18°) to (163°,28°) [d] |
| FS-2011-003 | 9.5 | 411, 197(121) | 35°, -23° | 35°, -22° | 10.5 (2.8) | AR 11171 | (25°, -19°) |
| FS-2011-004 | 11 | 1289, 908 (276) | 187°, -17° | 182°, -17° | 24.0 (5.0) | LEAR 11176 | (200°, -13°) |
| FS-2011-005 | 10 | 959, 574 (268) | 164°, 25° | 158°, 27° | 30.5 (5.8) | AR 11180 | (162°, 26°) |
| FS-2011-006 | 8 | 671, 425 (190) | 92°, 12° | 92°, 12° | 17.5 (6.5) | Remnant AR 11166 | (92°, 10°) |
| FS-2011-007 | 4 | 428, 158 (157) | 333°, 17° | 333°, 13° | 13.6 (3.4) | AR 11190 | (339°, 12°) |
| FS-2011-008 | 10 | 1341, 535 (416) | 26°, 22° | 28°, 19° | 23.6 (9.7) | AR 11205 | (34°, 14°) |
| FS-2011-009 | 11 | 880, 438 (265) | 340°, 13° | 338°, 16° | 17.3 (5.2) | AR 11208 (11190) | (318°, 13°) |
| FS-2011-010 | 10 | 648, 267 (221) | 269°, 17° | 265°, 19° | 10.1 (3.1) | Remnant AR 11193 | (270°, 16°) |
| FS-2011-011 | 4 | 530, 355 (160) | 333°, 14° | 332°, 9° | 15.9 (3.7) | Remnant 11208 | (325°, 9°) |
| FS-2011-012 | 10 | 1371, 1041 (293) | 165°, 18° | 159°, 20° | 17.7 (5.0) | LEAR 11236 | (174°, 17°) |

[a] In units of millionth hemisphere radian.
[b] Normalized to $C_m A_0$. (see text)
[c] Coordinates on date of NOAA designation for east limb ARs. For remnant ARs, west limb coordinates.
[d] Moves from AR 11163 to AR11164

**Table 4.  2012 Farside Seismic Regions and NOAA AR Associations**

| FS in 2012 | days detected seismically | Seismic Signature Strength S [a] – max, mean (standard deviation) | FS Start Coordinates (Carrington Longitude, Latitude) | FS End Coordinates (Carrington Longitude, Latitude) | EUV intensity $I$ [b] over time of FS, mean (stdev) | NOAA AR association? | NOAA AR Coordinates (Carrington Longitude, Latitude) [c] |
|---|---|---|---|---|---|---|---|
| 2011-53 | >5 [d] | 1228, 690 (341) | 306°, 24° | 308°, 20° | 13.1 (5.0) | LEAR 11395 | (304°, 16°) |
| 2012-01 | 4 | 1003, 666 (324) | 211°, 31° | 205°, 31° | 10.1 (0.8) | LEAR 11402 | (212°, 30°) |
| 2012-02 | 7.5 | 1920, 1222 (430) | 66°, 21° | 46°, 24° | 21.7 (3.0) | LEAR 11410 | (67°, 24°) |
| 2012-03 | 6 | 562, 408 (124) | 125°, 9° | 124°, 6° | 7.0 (1.1) | LEAR 11408 | (125°, 5°) |
| 2012-04 | 4 | 513, 206 (203) | 67°, -28° | 74°, -25° | 14.9 (5.1) | Merged remnants of AR11389 and AR11388 | (86°, -20°) |
| 2012-05 | 7.5 | 738, 366 (263) | 90°, -23° | 86°, -17° | 11.5 (7.7) | | (101°, -26°) |
| 2012-04 & 05 combined | 8 | 748, 377 (252) | | | 11.5 (7.7) | Merged remnants of AR11389 and AR11388 | |
| 2012-06 | 7 | 914, 585 (277) | 206°, 31° | 204°, 28° | 25.0 (2.5) | AR11419 [e] | (203°, 28°) |
| 2012-07 | 8.5 | 437, 311 (90) | 56°, 18° | 52°, 20° | 14.7 (5.5) | LEAR 11423 | (56°, 18°) |
| 2012-08 | 3.5 | 515, 425 (107) | 30°, 11° | 29°, 11° | 13.0 (1.3) | AR11424 [Remnant AR 11415] | (45°, 9°) [(24°, 10°)] |
| 2012-09 | 11 | 852, 478 (183) | 302°, 19° | 297°, 19° | 11.7 (4.3) | NOAA plage 11451 | (308°,17°) |
| 2012-10 | 2.5 | 538, 361 (192) | 152°, -4° | 149°, -2° | 3.3 (0.2) | None [f] | |

[a] **In units of millionth hemisphere radian.**
[b] **Normalized to $C_m A_0$ (see text)**
[c] Coordinates on date of NOAA designation for east limb ARs. For remnant ARs, west limb coordinates.
[d] Detection begins outside analysis time window
[e] No seismic detection of nearby LEAR AR11420
[f] Small EUV bright region at this location

**Figures & Captions**

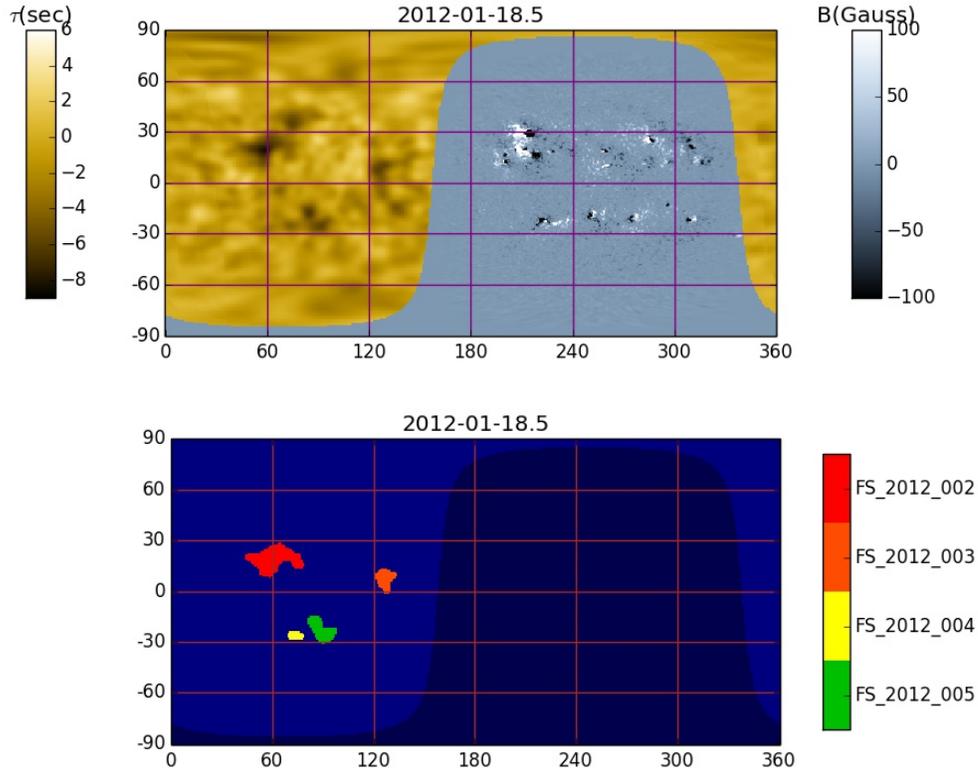

**Figure 1** Top: Composite seismic Carrington map (0 to 360° in longitude, -90° to +90° in latitude) for 2012 January 18 at 12UT produced using the 2014 five-day processing. The far side portion of the map shows the seismic signal τ on a logarithmic scale, where τ is the wave travel time shift; the Earth side region shows the HMI magnetogram (magnetic field B is in Gauss). Bottom: Four far-side strong field regions that were identified from the seismic signatures in the composite map above. Detected far-side (FS) regions are labeled by year and number, e.g., FS-2012-002 was the second strong field region to be detected seismically on the far side in 2012.

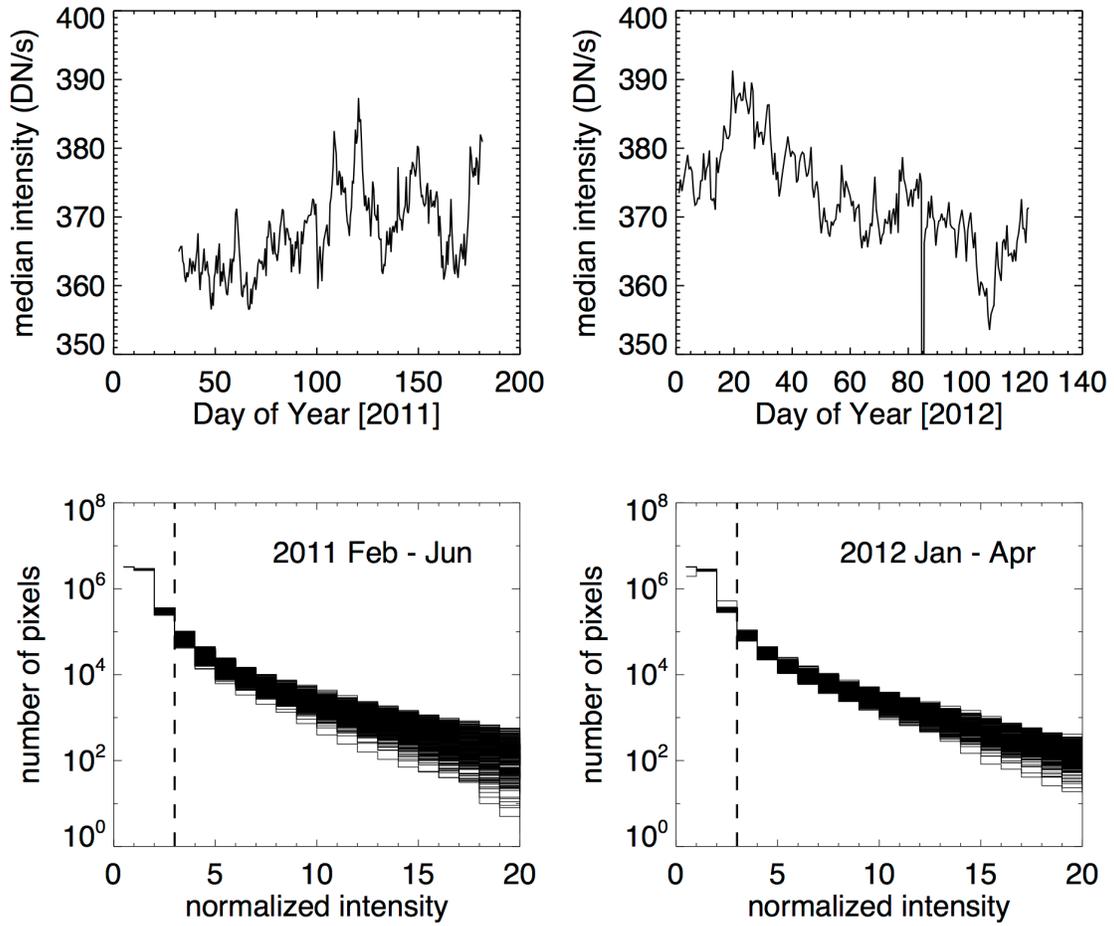

**Figure 2** Top: median intensity (in DN/s) of each EUV 304 Å map for 2011 (left) and 2012 (right). Bottom: histogram of pixel intensities normalized to the median for all EUV 304 Å maps in 2011 (left) and 2012 (right). The vertical dashed lines indicate the level of 3 times the median, which is the threshold for the definition of plage pixels used in this paper.

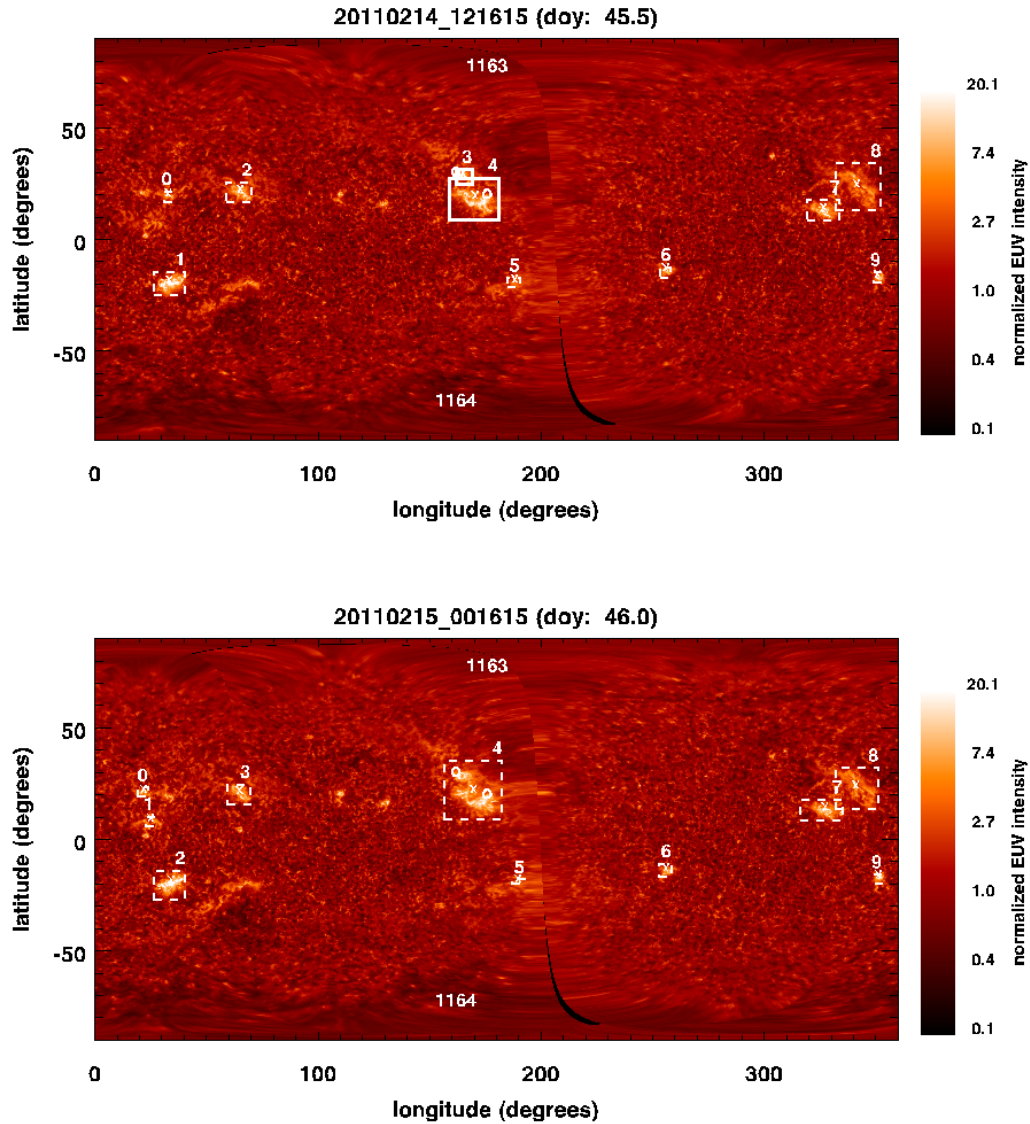

**Figure 3** Examples of plage identification and association with NOAA active regions for two maps, 2011 February 14 at 12 UT (top) and 2011 February 15 at 0 UT (bottom). White boxes mark all automatically identified plages in each map, and the centroids of the plages are marked with an "x". The location of an NOAA active region is marked with "o", and the NOAA active region number is given at the top or bottom of the map at its longitude. For these two days, AR 11163 and AR 11164 are marked. A solid white box marks a plage that is associated with a NOAA active region. The color scale shows the intensity of the EUV map normalized to the median of the map.

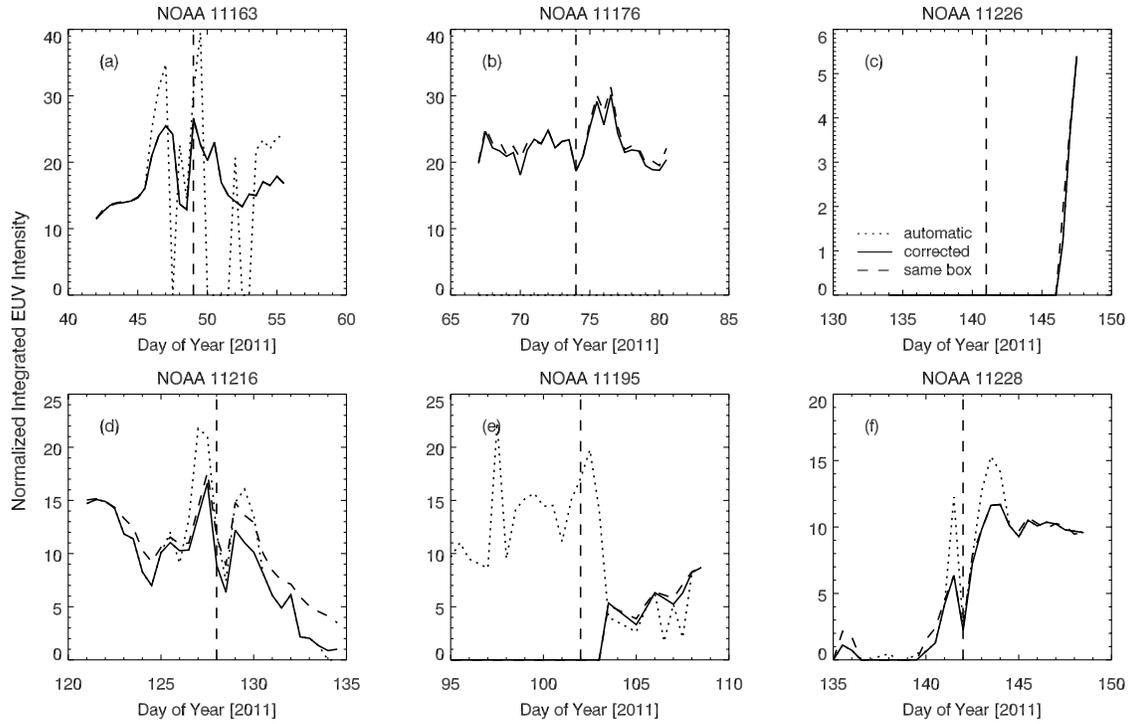

**Figure 4** Time history of the integrated EUV intensity (normalized to the quiet Sun intensity integrated over an area of 0.02 steradian) of several plages that have been associated with LEARs in 2011, calculated using three different methods. In each panel, EUV light curves are presented from automatically identified boxes (dotted line), corrected boxes (solid line), and a constant box (dashed line) during 14 days before appearance of the region at the Earth side. Vertical dashed lines indicate 7 days before Earth-side appearance.

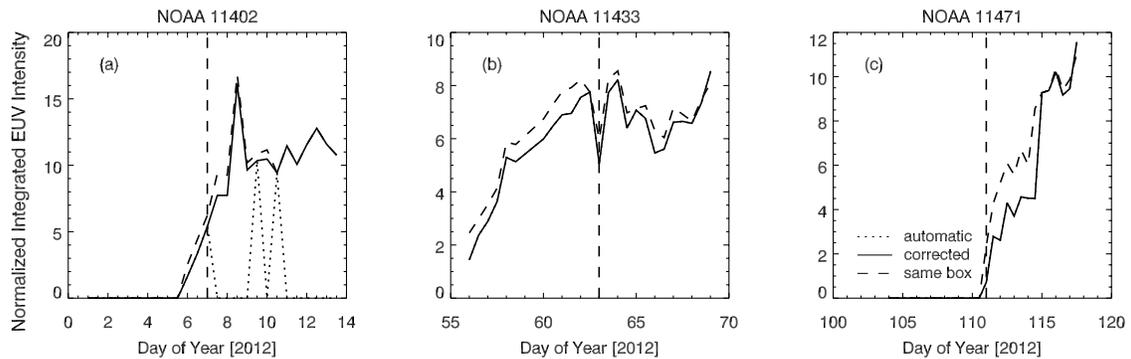

**Figure 5** Normalized Integrated EUV intensity of several plages associated with LEARs in 2012, calculated using three different methods. See caption to Figure 4.

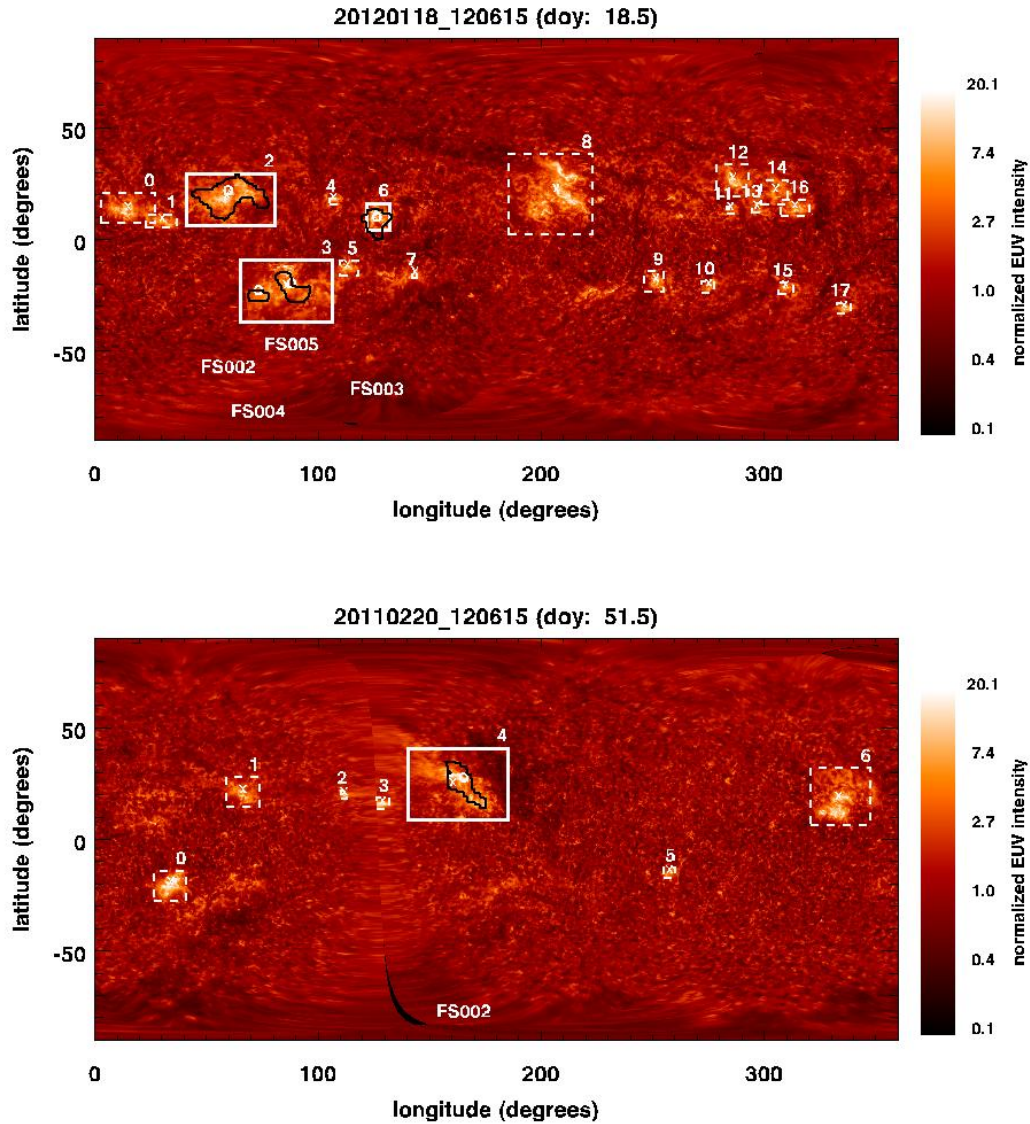

**Figure 6** Top: Four FS contours, from the lower map shown in Figure 1, overlaid on concurrent EUV map for 2012 January 18 at 12 UT showing correspondence of the seismic regions to EUV plages and their corresponding boxes. Centroids of the FS regions are marked with 'o', the FS label is at the longitude of the centroid. The centroids of EUV plage boxes are marked with 'x'. Bottom: EUV map for 2011 February 20 at 12 UT showing an EUV plage association with a far-side seismic (FS) signal overlaid on the EUV map. Here the large plage (box 4) encompasses a region that includes two future NOAA active regions, AR 11163 and 11164. See text **for further discussion.**

.

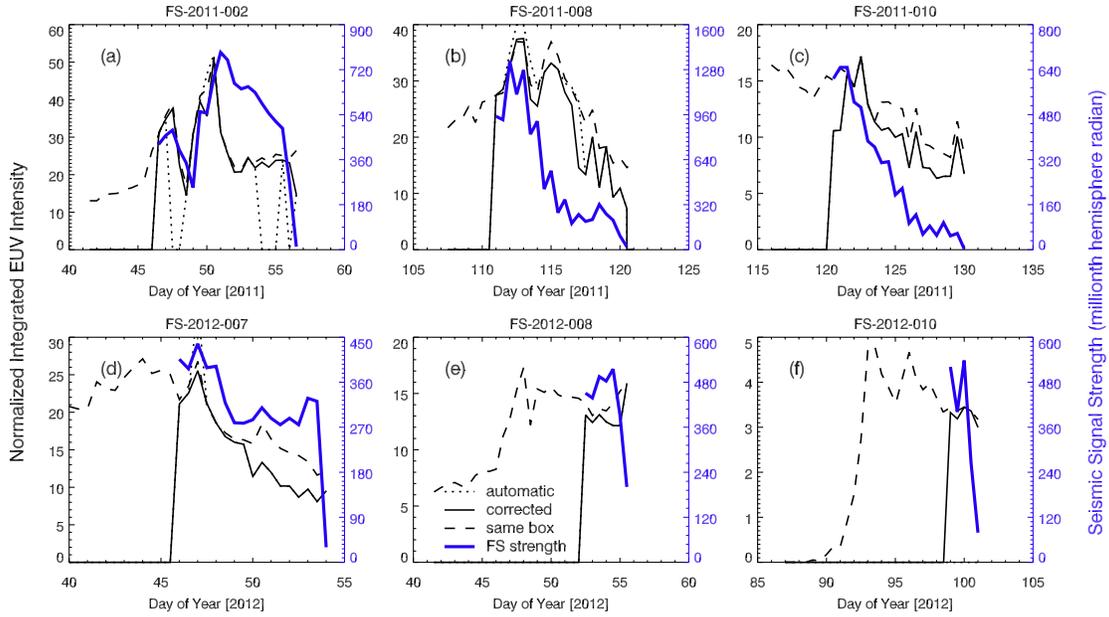

**Figure 7** Time evolution of the integrated EUV intensity of six plages associated with far-side seismic regions. The EUV intensity (normalized to the quiet Sun intensity integrated over an area of 0.02 steradian) is calculated in three ways, as in Figures 4 and 5. The corresponding seismic signal strength S (in units of millionth of hemisphere radian; scale on the right axis), is shown as a thick blue line, plotted during the time of the seismic detection.

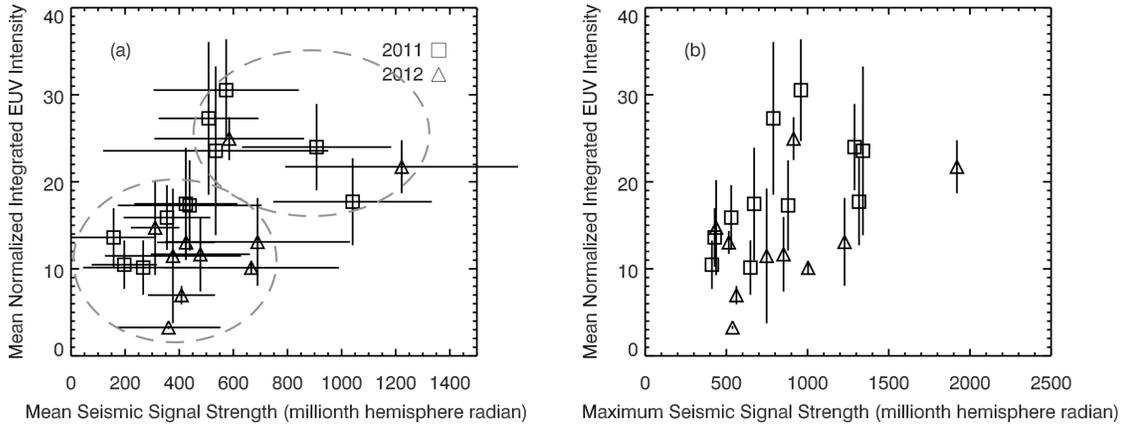

**Figure 8** (a) Comparison of the mean integrated EUV intensity ⟨*I*⟩ (normalized to the quiet Sun intensity times the spherical area 0.02 steradian) and mean seismic signal strength ⟨*S*⟩ (in units of a millionth of a hemisphere radian) for the associated regions, both averaged over the time of the seismic detections. (b) Comparison of the mean integrated EUV intensity ⟨*I*⟩, as in (a), and the maximum (over time of detection) seismic signal strength S. The vertical and/or horizontal bars indicate the standard deviation of the EUV intensity and/or seismic strength.

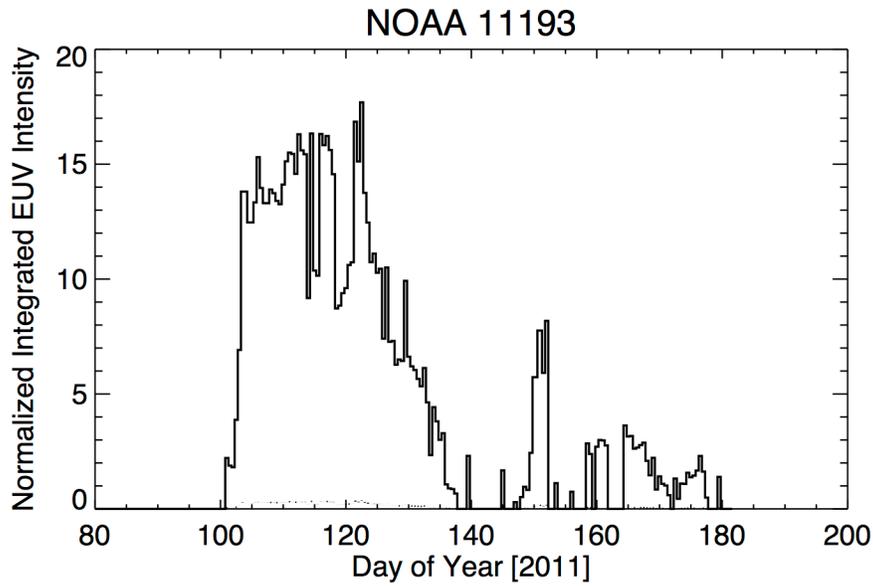

**Figure 9** Integrated EUV intensity in dynamically and automatically determined boxes, which are associated with the active region NOAA11193, during the life time of the active region. The evolution is consistent with the EUV intensity measured by Ugarte-Urra et al. (2015) although they used a very different method.